\documentclass[aps,pra,twocolumn,superscriptaddress]{revtex4}
\usepackage{amssymb}
\usepackage{amsmath}
\usepackage{graphicx}
\usepackage{epsfig}
\usepackage{subfigure}
\usepackage{color}

\begin{document}

\title{Optomechanical transistor with mechanical gain}
\author{X. Z. Zhang}
\affiliation{Beijing Computational Science Research Center, Beijing, 100193, China}
\affiliation{College of Physics and Materials Science, Tianjin Normal University, Tianjin
300387, China}
\author{Lin Tian}
\affiliation{University of California, Merced, 5200 North Lake Road, Merced, California
95343, USA}
\author{Yong Li}
\affiliation{Beijing Computational Science Research Center, Beijing, 100193, China}
\affiliation{Synergetic Innovation Center of Quantum Information and Quantum Physics,
University of Science and Technology of China, Hefei, Anhui 230026, China}
\affiliation{Synergetic Innovation Center for Quantum Effects and Applications, Hunan
Normal University, Changsha 410081, China}
\date{\today }

\begin{abstract}
We study an optomechanical transistor, where an input field can be
transferred and amplified unidirectionally in a cyclic three-mode
optomechanical system. In this system, the mechanical resonator is coupled
simultaneously to two cavity modes. We show that it only requires a finite
mechanical gain to achieve the nonreciprocal amplification. Here the
nonreciprocity is caused by the phase difference between the linearized
optomechanical couplings that breaks the time-reversal symmetry of this
system. The amplification arises from the mechanical gain, which provides an
effective phonon bath that pumps the mechanical mode coherently. This effect
is analogous to the stimulated emission of atoms, where the probe field can
be amplified when its frequency is in resonance with that of the anti-Stokes
transition. We show that by choosing optimal parameters, this optomechanical
transistor can reach perfect unidirectionality accompanied with strong
amplification. In addition, the presence of the mechanical gain can result
in ultra-long delay in the phase of the probe field, which provides an
alternative to controlling light transport in optomechanical systems.
\end{abstract}

\maketitle


%

\section{Introduction}

\label{Introduction}

The interaction between light and mechanical objects in the low-energy scale
has been intensively studied both in theory and in experiment during the
past two decades. Given the rapid advance in microfabrication~\cite%
{Aspelmeyer, Meystre, Metcalfe}, cavity optomechanical systems have been
exploited for both fundamental questions and various applications. Such
systems provide an appealing platform to study the quantum behavior of
macroscopic system~\cite{Vitali}. Meanwhile, applications of optomechanical
systems, such as ultra-sensitive measurement in the molecular scale~\cite%
{Rugar, Krause, Regal, Teufel,Forstner,Xu}, weak-force detection~\cite%
{Arvanitaki}, quantum wavelength conversion between microwave and optical
frequencies~\cite{Mancini, TianAnnPhys}, and quantum illumination~\cite%
{Barzanjeh}, have been investigated. Furthermore, optomachanical systems
have also been used to demonstrate quantum optical effects, such as
optomechanically induced transparency and absorption ~\cite{OMIT1,
OMIT2,OMIT3, OMIT4, OMIT5, OMIT6, OMslowlight1, OMslowlight2, OMIA1,OMIA2}
and optomechanically induced amplification~\cite%
{OMAmplification1,OMAmplification2}.

Among these applications, nonreciprocal transmission and amplification of
light fields are of great interest, similar to their analogues in electronic
devices. The nonreciprocal devices, which exhibit asymmetric response if the
input and output channels are interchanged, can protect unwanted singles
from entering into the network, where are essential to signal processing and
communications.
At the heart of the nonreciprocal devices is an element that breaks the
Lorentz reciprocity of the system~\cite{Baets}. Effects that have been used
to realize the nonreciprocity include the magneto-optical Faraday effect in
ferrite materials~\cite{Auld,Milano,Fay, Bi}, parametric modulation of
system parameters~\cite%
{ParametricModulation1,ParametricModulation2,ParametricModulation3,ParametricModulation4}%
, optical nonlinearity~\cite{Chang,Tang}, chiral light-matter interaction~%
\cite{Chiral}, and the rotation of device in the real space~\cite{Fleury}.
It has been shown that the nonreciprocal propagation of light can be
realized with optical devices~\cite{Manipatruni, Hafezi, shen, Kim}.
Meanwhile, unconventional propagation of light has been demonstrated by
engineering effective non-Hermitian Hamiltonians in optical systems~\cite%
{Guo, K. G. Makris, Z. H. Musslimani, S. Klaiman, S. Longhi, Zheng, Graefe,
Miroshnichenko}, which can be used to realize on-chip isolators and
circulators~\cite{Hamidreza}. Recently, $\mathcal{PT}$ symmetry breaking in
optomechanical systems with coupled cavities, often accompanied by the
coalescence of eigenstates at an exceptional point in the discrete spectrum,
has been studied~\cite{JingH, LiuYL}, and low-power phonon emissions~\cite%
{JingH}, chaos~\cite{LXY}, non-reciprocal energy transfer~\cite{XuH}, and
asymmetric mode switching~\cite{Doppler} have been observed. More recently,
optomechanical isolators, circulators, and directional amplifiers have been
studied in multi-mode systems by modulating the gauge-invariant phases~\cite%
{Fang,XXW,XXW2,Tian,Metelmann,YLZhang, Ranzani, Peterson, Malz,
Ruesink,Mohammad,Bernier}.

Here we present a scheme for realizing an optomechanical transistor in a
cyclic three-mode optomechanical system with finite mechanical gain. In this
system, two optical modes are linearly coupled with each other and are also
coupled simultaneously to a common mechanical mode. The phase difference
between the optomechanical couplings breaks the time-reversal symmetry of
this system and ensures nonreciprocity in the state transmission. Meanwhile,
amplification arises from the mechanical gain, which induces a phonon-photon
parametric process. Compared to our previous work~\cite{liyong}, this
approach does not require the frequency matching between the pump fields on
the cavity and the mechanical modes. Furthermore, we show that within the
operational parameter window of the optomechanical transistor, an ultra-long
delay in the phase of the probe field occurs due to the finite mechanical
gain. These findings provide an alternative way to achieving controlled
light transport in optomechanical systems and can stimulate future works in
light amplification with optomechanical devices.

This paper is organized as follows. In Sec.~\ref{model}, we introduce the
three-mode optomechanical system with finite mechanical gain. The stability
of this system is also discussed in this section. We then derive the
transmission coefficients of this system in a generic setting in Sec.~\ref%
{sec3}. The behavior of the optomechanical transistor and the ultra-long
delay in the phase of the probe field are studied in detail in Sec.~ \ref%
{optical response}, Finally, conclusions are given in Sec.~\ref{conclusions}.

\section{The model}

\label{model}

\begin{figure}[tbp]
\centering
\includegraphics[bb=48 458 551 692, width=0.45\textwidth, clip]{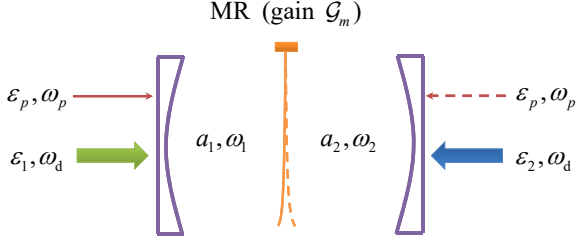} %
\caption{The schematic of a cyclic three-mode optomechanical system driven
by two pump fields of amplitudes $\protect\varepsilon_{1}$ and $\protect%
\varepsilon_{2}$ with frequency $\protect\omega _{d}$. A probe field with
amplitude $\protect\varepsilon_{p}$ and frequency $\protect\omega _{p}$ is
applied to one of the cavities (to cavity 1 from the left hand side or
cavity 2 from the right hand side). A mechanical gain $\protect\gamma _{m}$
is engineered on the mechanical mode with frequency $\protect\omega _{m}$.
The cavities and the mechanical resonator are coupled via radiation-pressure
forces and the cavities are directly coupled to each other.}
\label{fig1}
\end{figure}

Consider an optomechanical system that contains a mechanical mode with
frequency $\omega _{m}$ and two cavity modes with frequencies $\omega _{1}$
and $\omega _{2}$, respectively, as illustrated in Fig.~\ref{fig1}. The
Hamiltonian of this system has the form ($\hbar =1$)
\begin{eqnarray}
H &=& H_{0}+H_{I}+H_{d},  \label{H} \\
H_{0} &=& \omega _{1}a_{1}^{\dag }a_{1}+\omega _{2}a_{2}^{\dag }a_{2}+\omega
_{m}b^{\dag }b, \\
H_{I}& =& J(a_{1}^{\dag }a_{2}+a_{1}a_{2}^{\dag }) +\sum_{i}g_{i}a_{i}^{\dag
}a_{i}( b+b^{\dag }), \\
H_{d}& =&\sum_{i}i\varepsilon _{i}( a_{i}^{\dag }e^{-i\omega
_{d}t}e^{i\theta_{i}}-\text{H.c.}).
\end{eqnarray}
Here $H_{0}$ is the Hamiltonian of the uncoupled cavity and mechanical
modes, where $a_{i}^{\dag }$ ($a_{i}$) for $i=1,2$ and $b^{\dag }$ ($b$) are
the corresponding creation (annihilation) operators of these modes. The
Hamiltonian $H_{I}$ describes the linear interaction between the cavities
with coupling strength $J$ and the radiation-pressure interactions between
the cavity and the mechanical modes with coupling strength $g_{1}$ and $g_{2}
$. The Hamiltonian $H_{d}$ represents the pump fields applied to the
cavities with frequency $\omega_{d}$, amplitudes $\varepsilon_{1,2}$ and
phases $\theta _{1,2}$. Without loss of generality, we assume that the
parameters $J$, $g_{1,2}$, and $\varepsilon _{1,2}$ are real numbers. In the
rotating frame of $\omega _{d}$, the Hamiltonian becomes
\begin{align}
H_{\text{rot}}&=\sum_{i}\Delta _{i}a_{i}^{\dag }a_{i}+\omega_{m}b^{\dag
}b+J( a_{1}^{\dag }a_{2}+a_{1}a_{2}^{\dag })  \notag \\
& +\sum_{i}g_{i}a_{i}^{\dag }a_{i}\left( b+b^{\dag }\right) + i\varepsilon
_{i}( a_{i}^{\dag }e^{i\theta _{i}}-\text{H.c.}),  \label{Hrot}
\end{align}
where $\Delta _{i}=\omega _{i}-\omega _{d}$ ($i=1,2$) is the detuning of the
pump field from the cavity resonance.

We assume the cavity and the mechanical modes are subject to input noise
denoted by $f_{i}^{in}$ ($i=1,2 $) for the cavity input operators and $%
f_{b}^{in}$ for the mechanical input with $\langle f_{i}^{in}\rangle
=\langle f_{b}^{in}\rangle =0$. With Hamiltonian (\ref{Hrot}), the Quantum
Langevin equations (QLEs) for the above optomechanical system are
\begin{eqnarray}
\dot{a}_{1}& =&\left\{ -\gamma _{1}-i\left[ \Delta _{1}+g_{1}\left(
b+b^{\dag}\right) \right] \right\} a_{1}-iJa_{2}  \notag \\
& &+\varepsilon _{1}e^{i\theta _{_{1}}}+\sqrt{2\gamma _{1}}f_{1}^{in},
\label{NQE1} \\
\dot{a}_{2}& =&\left\{ -\gamma _{2}-i\left[ \Delta _{2}+g_{2}\left(
b+b^{\dag }\right) \right] \right\} a_{2}-iJa_{1}  \notag \\
& &+\varepsilon _{2}e^{i\theta _{_{2}}}+\sqrt{2\gamma _{2}}f_{2}^{in},
\label{NQE2} \\
\dot{b}& =&\left( \mathcal{G}_{m}-i\omega _{m}\right)
b-i\left(g_{1}a_{1}^{\dag }a_{1}+g_{2}a_{2}^{\dag }a_{2}\right)  \notag \\
&&+\sqrt{2\mathcal{G}_{m}}f_{b}^{in},  \label{NQE3}
\end{eqnarray}
where $\gamma_{i}$ $\left( i=1,2\right) $ is the decay rate of the
corresponding cavity mode and $\mathcal{G}_{m}$ denotes the controllable
gain of the mechanical mode. In practical systems, the mechanical gain can
be obtained with various methods, e.g., through phonon lasing or by coupling
the mechanical mode to another cavity mode and applying blue-detuned driving
to the cavity~\cite{LiuYL}.

\begin{figure*}[tbp]
\centering
\includegraphics[bb=16 20 400 424, width=0.4\textwidth,
clip]{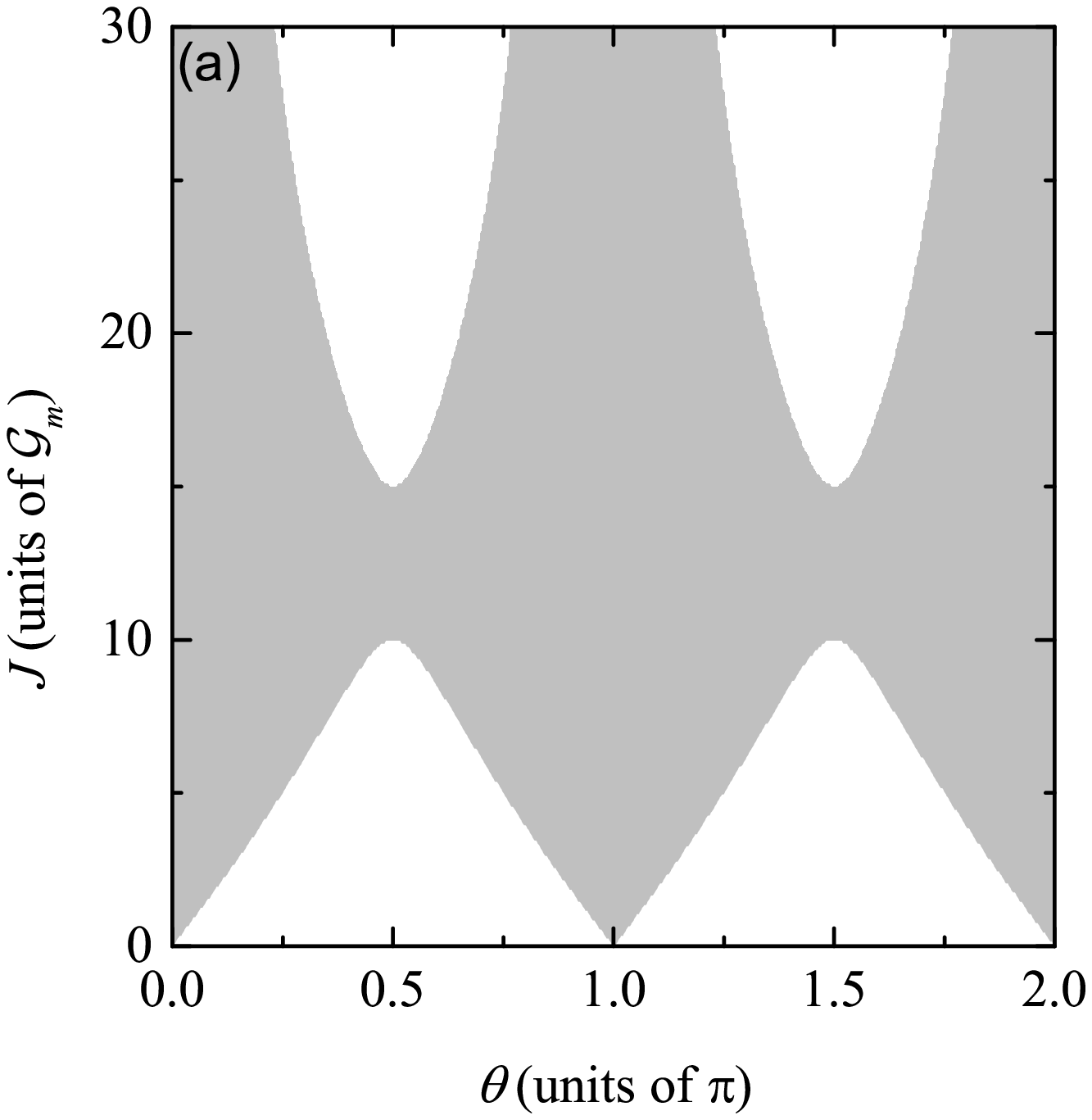}
\includegraphics[bb=16 20 400 424, width=0.4\textwidth,
clip]{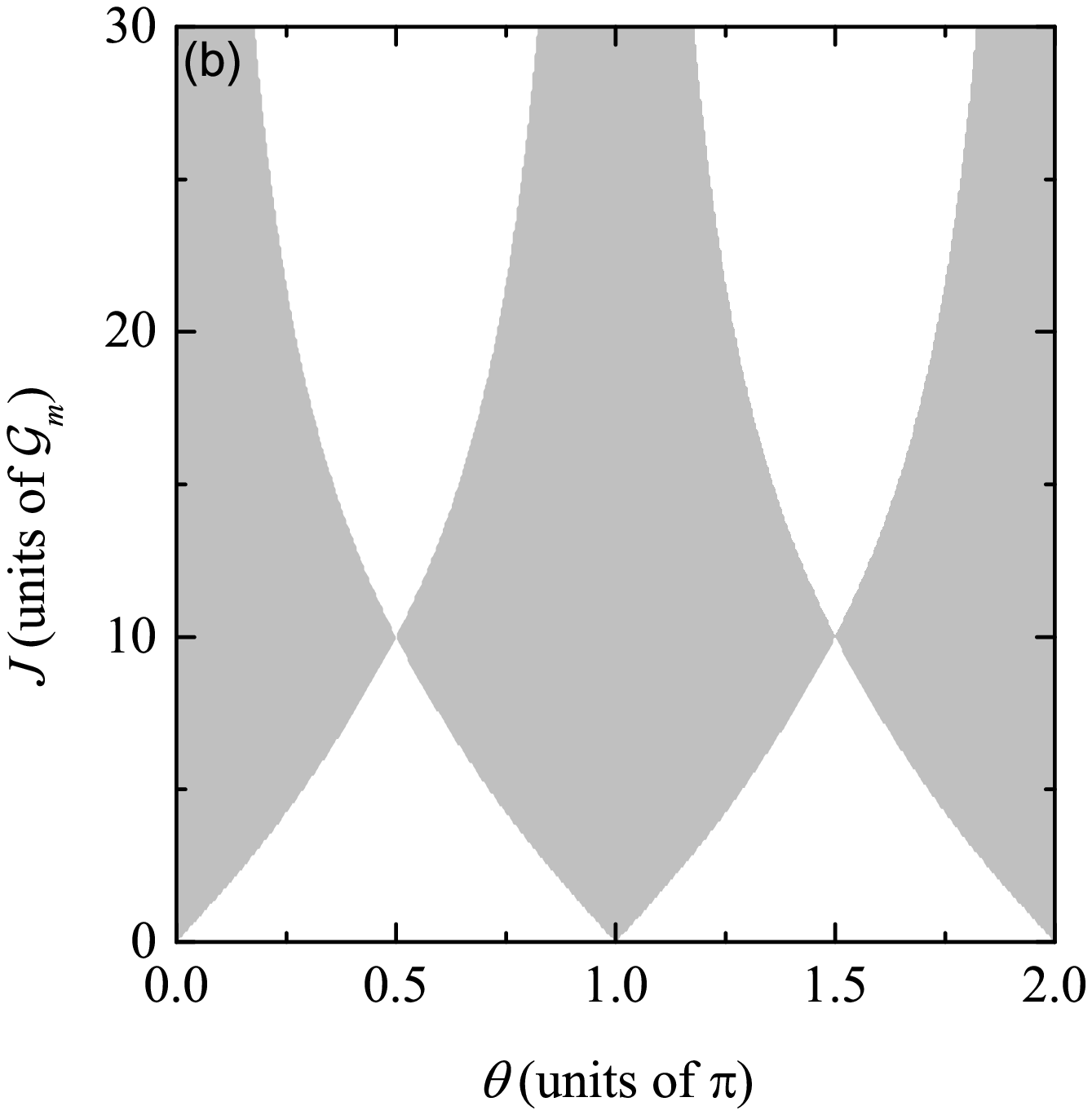}
\caption{(Color online) Numerical calculation of the stability of this
system with the parameters (a) $\protect\gamma _{1}=10\mathcal{G}_{m}$, $%
\protect\gamma _{2}=15\mathcal{G}_{m}$ and (b) $\protect\gamma _{1}=10%
\mathcal{G}_{m} $, $\protect\gamma _{2}=10\mathcal{G}_{m}$. Other parameters
are $G_{1} =\left\vert G_{2}\right\vert \equiv G=\protect\sqrt{J\mathcal{G}%
_{m}/\sin \protect\theta }$, and $\protect\omega _{m}/\mathcal{G}_{m}=10^{3}$%
. Each panel contains two regions. The gray (white) regions represent the
stable (unstable) regions of this system. In particular, in (b), when $J=10\,%
\mathcal{G}_{m}$, the system is stable with all values $\protect\theta $
except for $\protect\theta =\protect\pi /2$.}
\label{fig2}
\end{figure*}

With strong pumping, the steady-state solutions of the cavity modes $\langle
a_{i}\rangle $ and of the mechanical mode $\langle b\rangle $ can be
obtained as
\begin{eqnarray}
\left\langle a_{1}\right\rangle &=&\frac{\left( \gamma _{2}+i\Delta
_{2}^{\prime }\right) \varepsilon _{1}e^{i\theta _{1}}-iJ\varepsilon
_{2}e^{i\theta _{2}}}{\left( \gamma _{1}+i\Delta _{1}^{\prime }\right)
\left( \gamma _{2}+i\Delta _{2}^{\prime }\right) +J^{2}}, \\
\left\langle a_{2}\right\rangle &=&\frac{\left( \gamma _{1}+i\Delta
_{1}^{\prime }\right) \varepsilon _{2}e^{i\theta _{2}}-iJ\varepsilon
_{1}e^{i\theta _{1}}}{\left( \gamma _{1}+i\Delta _{1}^{\prime }\right)
\left( \gamma _{2}+i\Delta _{2}^{\prime }\right) +J^{2}}, \\
\left\langle b\right\rangle &=&\frac{-i(g_{1}\left\vert \left\langle
a_{1}\right\rangle \right\vert ^{2}+g_{2}\left\vert \left\langle
a_{2}\right\rangle \right\vert ^{2})}{-\mathcal{G}_{m}+i\omega _{m}}
\end{eqnarray}%
with $\Delta _{i}^{\prime }=\Delta _{i}+g_{i}[\langle b\rangle +\langle
b\rangle ^{\ast }]$. These coupled equations can be solved
self-consistently. By assuming each operator is a sum of the steady-state
solution and its quantum fluctuation, i.e., $a_{i}=\langle a_{i}\rangle
+\delta a_{i}$ and $b=\langle b\rangle +\delta b$, and neglecting the
nonlinear terms, we obtain a set of linearized QLEs:
\begin{eqnarray}
\delta \dot{a}_{1} &=&\left( -\gamma _{1}-i\Delta _{1}^{\prime }\right)
\delta a_{1}-iG_{1}\left( \delta b+\delta b^{\dag }\right)  \notag \\
&&-iJ\delta a_{2}+\sqrt{2\gamma _{1}}f_{1}^{in},  \label{QLE1} \\
\delta \dot{a}_{2} &=&\left( -\gamma _{2}-i\Delta _{2}^{\prime }\right)
\delta a_{2}-iG_{2}\left( \delta b+\delta b^{\dag }\right)  \notag \\
&&-iJ\delta a_{1}+\sqrt{2\gamma _{2}}f_{2}^{in},  \label{QLE2} \\
\delta \dot{b} &=&\left( \mathcal{G}_{m}-i\omega _{m}\right) \delta
b-i(G_{1}\delta a_{1}^{\dag }+G_{1}^{\ast }\delta a_{1})  \notag \\
&&-i(G_{2}\delta a_{2}^{\dag }+G_{2}^{\ast }\delta a_{2})+\sqrt{2\mathcal{G}%
_{m}}f_{b}^{in},  \label{QLE3}
\end{eqnarray}%
where $G_{i}=g_{i}\langle a_{i}\rangle $ ($i=1,2$) is the effective linear
coupling between the $i$th cavity and the mechanical mode. We assume that
the system is operated in the resolved sideband regime with $\gamma _{i},%
\mathcal{G}_{m},G_{i}\ll \omega _{m}$ and $\Delta _{i}^{\prime }\sim \omega
_{m}$. With these assumptions, we can apply the rotating-wave approximation
to the above QLEs and neglect the fast-oscillating counter-rotating terms.
The QLEs become
\begin{eqnarray}
\delta \dot{a}_{1} &=&-\Gamma _{10}\delta a_{1}-iG_{1}\delta b-iJ\delta
a_{2}+\sqrt{2\gamma _{1}}f_{1}^{in},  \label{a1QLE} \\
\delta \dot{a}_{2} &=&-\Gamma _{20}\delta a_{2}-iG_{2}\delta b-iJ\delta
a_{1}+\sqrt{2\gamma _{2}}f_{2}^{in},  \label{a2QLE} \\
\delta \dot{b} &=&-\Gamma _{m0}\delta b-iG_{1}^{\ast }\delta
a_{1}-iG_{2}^{\ast }\delta a_{2}+\sqrt{2\mathcal{G}_{m}}f_{b}^{in},
\label{bQLE}
\end{eqnarray}%
where $\Gamma _{i0}=\gamma _{i}+i\Delta _{i}^{\prime }$, and $\Gamma _{m0}=-%
\mathcal{G}_{m}+i\omega _{m}$. For simplicity, we rewrite the linearized
QLEs in matrix form with
\begin{equation}
\frac{d}{dt}\lambda =-M\lambda +\Upsilon \lambda ^{in},
\end{equation}%
where the fluctuation vector $\lambda =(\delta a_{1},\delta a_{2},\delta
b)^{T}$, the input field $\lambda ^{in}=(f_{1}^{in},\text{ }f_{2}^{in},\text{
}f_{b}^{in})^{T}$, the coupling matrix for the input operators $\Upsilon =$
diag$\left( \sqrt{2\gamma _{1}},\sqrt{2\gamma _{2}},\sqrt{2\mathcal{G}_{m}}%
\right) $, and dynamic matrix
\begin{equation}
M=\left(
\begin{array}{ccc}
\Gamma _{10} & iJ & iG_{1} \\
iJ & \Gamma _{20} & iG_{2} \\
iG_{1}^{\ast } & iG_{2}^{\ast } & \Gamma _{m0}%
\end{array}%
\right) .  \label{CM}
\end{equation}

The stability of this optomechanical system can be influenced by the
mechanical gain. The stability condition for this system can be derived
using the Routh-Hurwitz criterion, which is equivalent to the requirement
that the eigenvalues of matrix $M$ have no positive real part. In Fig.~\ref%
{fig2}, we plot two typical cases that are employed to investigate the
optical response of this system in the following sections, where the gray
regions are stable and the white regions are unstable. When $J=\gamma
_{1}=\gamma _{2}=10\mathcal{G}_{m}$, as shown in Fig.~\ref{fig2}(b), the
system is stable with all the possible values of $\theta $ except for $%
\theta =\pi /2$. However, when the system parameters are $J=11\mathcal{G}_{m}
$, $\gamma _{1}=10\mathcal{G}_{m}$, and $\gamma _{1}=15\mathcal{G}_{m}$, the
stable region covers all values of $\theta $, which can be seen in Fig.~\ref%
{fig2}(a).

\section{Transmission coefficients}

\label{sec3}

Apply a probe field to cavity $1$ in the form of $i(\varepsilon
_{p}a_{1}^{\dag }e^{-i\omega _{p}t}-\text{H.c.})$, as illustrated by the
thin solid arrow in Fig.~\ref{fig1}. The response to a probe field applied
to cavity $2$ (the thin dashed arrow in Fig.~\ref{fig1}) can be obtained by
exchanging the subscripts $1$ and $2$ in the following results. We assume
that the amplitude of the probe field $\varepsilon _{p}$ is much smaller
than that of the control field $\varepsilon _{1,2}$, and the steady-state
solutions of the operators $a_{1}$, $a_{2}$, $b$ will not be affected by the
probe field. Hence the only change in the QLEs is that one extra term $%
\varepsilon _{p}e^{-i\left( \omega _{p}-\omega _{d}\right) t}$ is added to (%
\ref{a1QLE}). To solve this set of linear QLEs, we use another interaction
picture by transforming $\delta a_{i}\rightarrow \delta a_{i}e^{-i\left(
\omega _{p}-\omega _{d}\right) t}$, $f_{i}^{in}\rightarrow
f_{i}^{in}e^{-i\left( \omega _{p}-\omega _{d}\right) t}$ ($i=1,2$), and $%
f_{b}^{in}\rightarrow f_{b}^{in}e^{-i\left( \omega _{p}-\omega _{d}\right) t}
$. The corresponding QLEs become
\begin{eqnarray}
\delta \dot{a}_{1} &=&-\Gamma _{1}\delta a_{1}-iG_{1}\delta b-iJ\delta
a_{2}+\varepsilon _{p}+\sqrt{2\gamma _{1}}f_{1}^{in},  \label{QQLE1} \\
\delta \dot{a}_{2} &=&-\Gamma _{2}\delta a_{2}-iG_{2}\delta b-iJ\delta a_{1}+%
\sqrt{2\gamma _{2}}f_{2}^{in},  \label{QQLE2} \\
\delta \dot{b} &=&-\Gamma _{m}\delta b-iG_{1}^{\ast }\delta
a_{1}-iG_{2}^{\ast }\delta a_{2}+\sqrt{2\mathcal{G}_{m}}f_{b}^{in},
\label{QQLE3}
\end{eqnarray}%
where $\Gamma _{i}=\gamma _{i}+i\Delta _{i}^{\prime \prime }$ and $\Gamma
_{m}=-\mathcal{G}_{m}+i\Delta _{m}$ with $\Delta _{i}^{\prime \prime
}=\Delta _{i}^{\prime }-\left( \omega _{p}-\omega _{d}\right) $ and $\Delta
_{m}=\omega _{m}-\left( \omega _{p}-\omega _{d}\right) $ being the detunings
in the new frame.

The optical response of this system to the probe field can be obtained by
solving the steady state of Eqs.~(\ref{QQLE1}--\ref{QQLE3}). By setting $%
\delta \dot{a}_{i}=\delta \dot{b} =0$ and neglecting the noise terms, we
obtain
\begin{align}
\left\langle \delta a_{1}\right\rangle & =\varepsilon _{p}(\Gamma _{2}\Gamma
_{m}+\left\vert G_{2}\right\vert ^{2})/D,  \label{delta_1} \\
\left\langle \delta a_{2}\right\rangle & =-\varepsilon _{p}\left(
G_{1}^{\ast }G_{2}+iJ\Gamma _{m}\right) /D,  \label{delta_2} \\
\left\langle \delta b\right\rangle & =-\varepsilon _{p}(iG_{1}^{\ast }\Gamma
_{2}+JG_{2}^{\ast })/D  \label{delta_b}
\end{align}%
with the denominator
\begin{eqnarray}
D &=&J^{2}\Gamma _{m}+\Gamma _{m}\Gamma _{1}\Gamma _{2}+\left( \Gamma
_{1}\left\vert G_{2}\right\vert ^{2}+\Gamma _{2}\left\vert G_{1}\right\vert
^{2}\right)  \notag \\
&&-iJ\left( G_{1}^{\ast }G_{2}+G_{1}G_{2}^{\ast }\right) .
\end{eqnarray}%
The amplitudes $\left\langle \delta a_{i}^{out}\right\rangle $ of the
experimentally accessible cavity output fields are related to the cavity
field $\left\langle \delta a_{i}\right\rangle $ by the input-output relation%
\begin{equation}
\left\langle \delta a_{i}^{out}\right\rangle +\left\langle \delta
a_{i}^{in}\right\rangle =\sqrt{2\gamma _{i}^{e}}\left\langle \delta
a_{i}\right\rangle ,\text{(}i=1,2\text{)}  \label{input-output}
\end{equation}%
where $\left\langle \delta a_{1}^{in}\right\rangle =\varepsilon _{p}/\sqrt{%
2\gamma _{1}^{e}}$, $\left\langle \delta a_{2}^{in}\right\rangle =0$, and $%
\gamma _{i}^{e}$ is the external damping rate that describes the coupling
between the cavity mode and the input field. We can write $\gamma
_{i}^{e}=\eta \gamma _{i}$ with $\eta $ being the ratio between the external
damping rate and the total damping rate. For the coupling parameter $\eta
\ll 1$, the cavity is under-coupled; and when $\eta \simeq 1$, the cavity is
over-coupled. The ratio $\eta $ can be continuously adjusted in experiments~%
\cite{Cai,Spillane}. In this work, we consider the cases of over-coupled
cavities with $\eta =1$ and neglect the cavity intrinsic dissipation.

Using Eqs. (\ref{delta_1}, \ref{delta_2}, \ref{input-output}), the
transmission coefficient $t_{21}\equiv \partial \left\langle \delta
a_{2}^{out}\right\rangle /\partial \left\langle \delta
a_{1}^{in}\right\rangle $ can be derived as
\begin{equation}
t_{21}=-\frac{2\sqrt{\gamma _{1}\gamma _{2}}\left( G_{1}^{\ast
}G_{2}+iJ\Gamma _{m}\right) }{D}.  \label{t_12}
\end{equation}%
By interchanging indices $1$ and $2$ in (\ref{t_12}), we find that
\begin{equation}
t_{12}=-\frac{2\sqrt{\gamma _{1}\gamma _{2}}\left( G_{2}^{\ast
}G_{1}+iJ\Gamma _{m}\right) }{D}.  \label{t_21}
\end{equation}%
From (\ref{t_12}, \ref{t_21}), we find that by manipulating the phase
difference between the optomechanical couplings $G_{1}$ and $G_{2}$,
nonreciprocal propagation of the probe field can be achieved, i.e., $%
|t_{12}/t_{21}|$ can be adjusted by varying the phase difference. This
effect can be understood through the effective Hamiltonian associated with (%
\ref{QQLE1}--\ref{QQLE3}),
\begin{eqnarray}
H_{\text{eff}} &=&\sum_{i}\Delta _{i}^{\prime \prime }\delta a_{i}^{\dag
}\delta a_{i}+\Delta _{m}\delta b^{\dag }\delta b  \notag \\
&&+\sum_{i}G_{i}\delta a_{i}^{\dag }\delta b+J\delta a_{1}^{\dag }\delta
a_{2}+\text{H.c.},
\end{eqnarray}%
which describes a typical three-mode cyclic system. The propagation of light
fields in such a system depends strongly on the interference between
different paths in the loop. A non-zero phase difference between the
couplings $G_{1}$ and $G_{2}$ can break the time reversal symmetry of this
system and gives rise to nonreciprocal optical response~\cite{xu}. Compared
to our previous work~\cite{liyong}, the advantage of this scheme is that it
does not require the matching of the pump frequencies between the optical
and mechanical fields to achieve nonreciprocal propagation of the probe
field. The mechanical gain can be viewed as a coherent bath that converts
the beam-splitter operation between the mechanical mode and the cavities
into phonon-photon parametric processes. We will discuss these points in
detail in the following section.

\section{Nonreciprocal amplification and optical delay}

\label{optical response}

In this section, we will investigate the properties of the transmission
coefficients under a special setup, i.e., when the system acts as an
optomechanical transistor. We will show the feasibility of achieving signal
amplification and nonreciprocity and study the delayed output response in
this three-mode optomechanical system.

\begin{figure}[tbp]
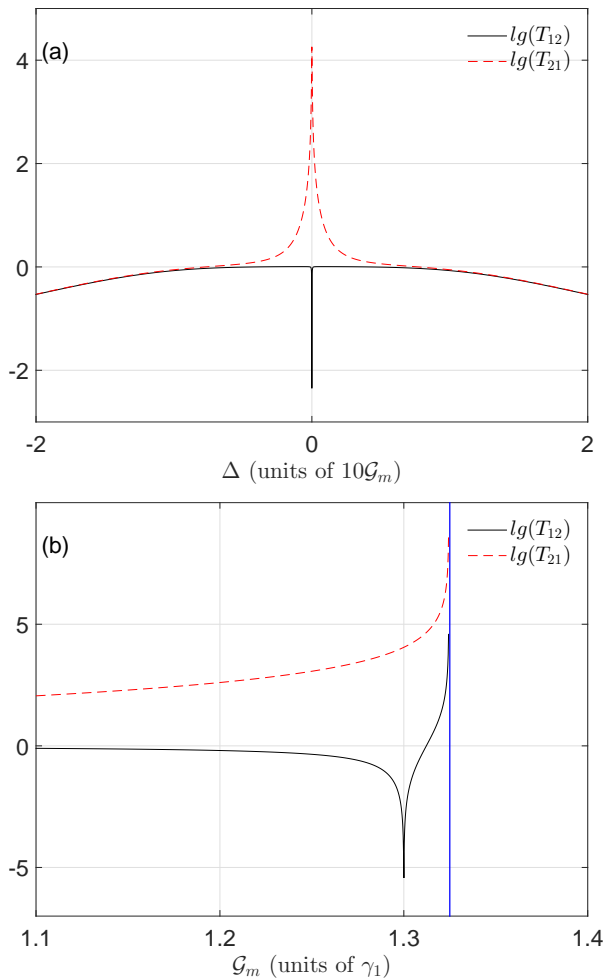

\centering
\includegraphics[bb=70 35 545 420, width=0.45\textwidth, clip]{fig3a.eps} %
\includegraphics[bb=70 35 545 420, width=0.45\textwidth, clip]{fig3b.eps}%
\caption{(Color online) (a) The logarithms of the transmission probabilities
$\lg {(T_{21})} $ and $\lg {(T_{22})}$ versus the detuning $\Delta $. Other
parameters are $\protect\gamma _{1}=10\mathcal{G}_{m}$, $\protect\gamma %
_{2}=15\mathcal{G}_{m}$, $J=11\mathcal{G}_{m}$, $\protect\theta =\protect\pi %
/2$, and $G_{1}=\left\vert G_{2}\right\vert \equiv G=\protect\sqrt{J\mathcal{%
G}_{m}}$. In the vicinity of $\Delta =0$, the transmission exhibits
unidirectional amplification in agreement with the analytical result. (b)
The logarithms of the transmission probabilities $\lg {(T_{12})}$ and $\lg {%
(T_{21})}$ versus the mechanical gain $\mathcal{G}_{m}$. Other parameters
are $\protect\gamma _{2}=1.5\protect\gamma _{1}$, $J=1.3\protect\gamma _{1}$%
, $\protect\theta =\protect\pi /2$, $G_{1}=\left\vert G_{2}\right\vert
\equiv G=J$, and $\Delta =0$. Here when $\mathcal{G}_{m}/\protect\gamma %
_{1}> 1.325$, the system becomes unstable.}
\label{fig3}
\end{figure}

\begin{figure*}[tbp]
\centering
\includegraphics[bb=16 20 471 517, width=0.40\textwidth,clip]{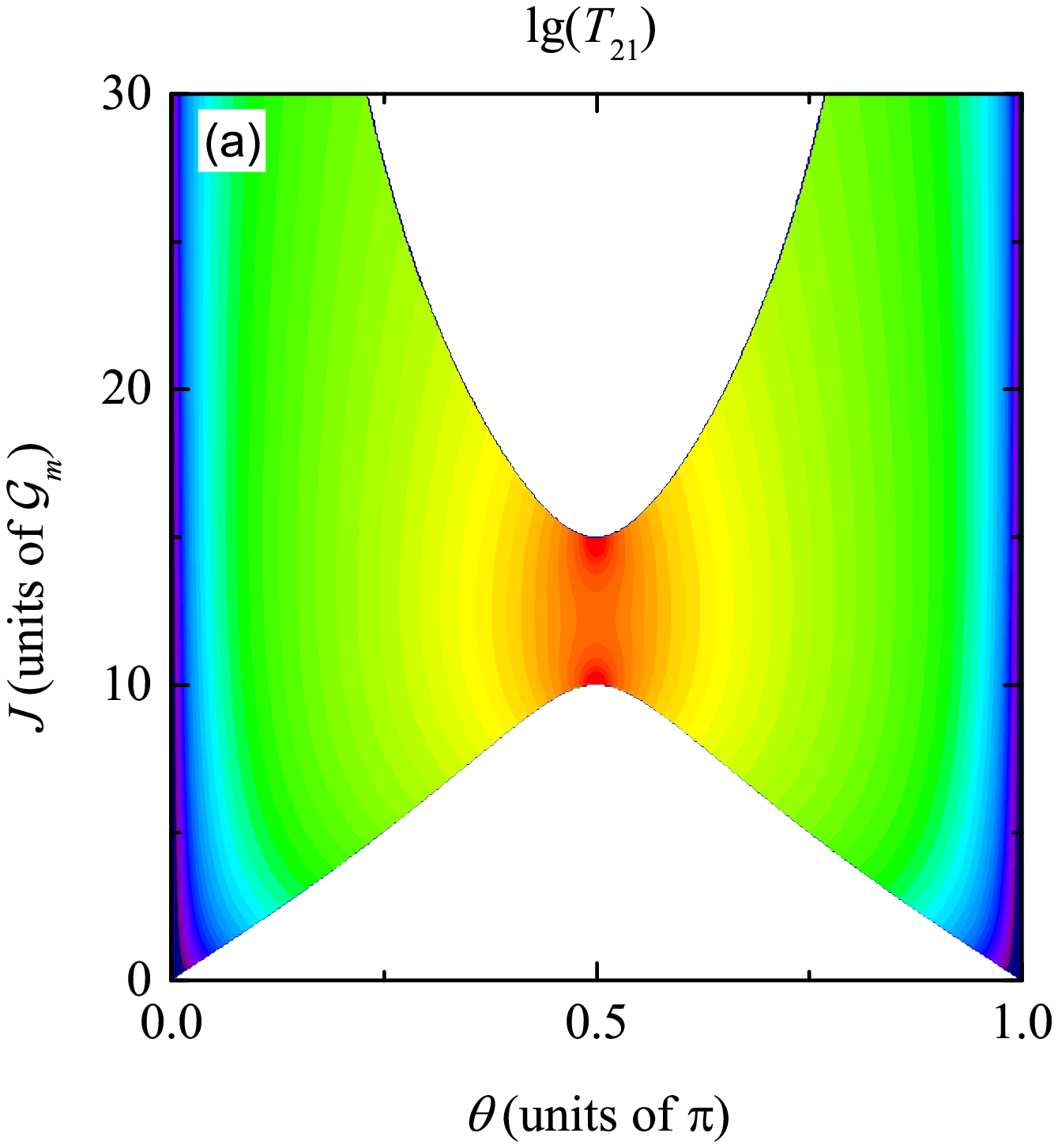} %
\includegraphics[bb=16 20 547 517, width=0.4675\textwidth,clip]{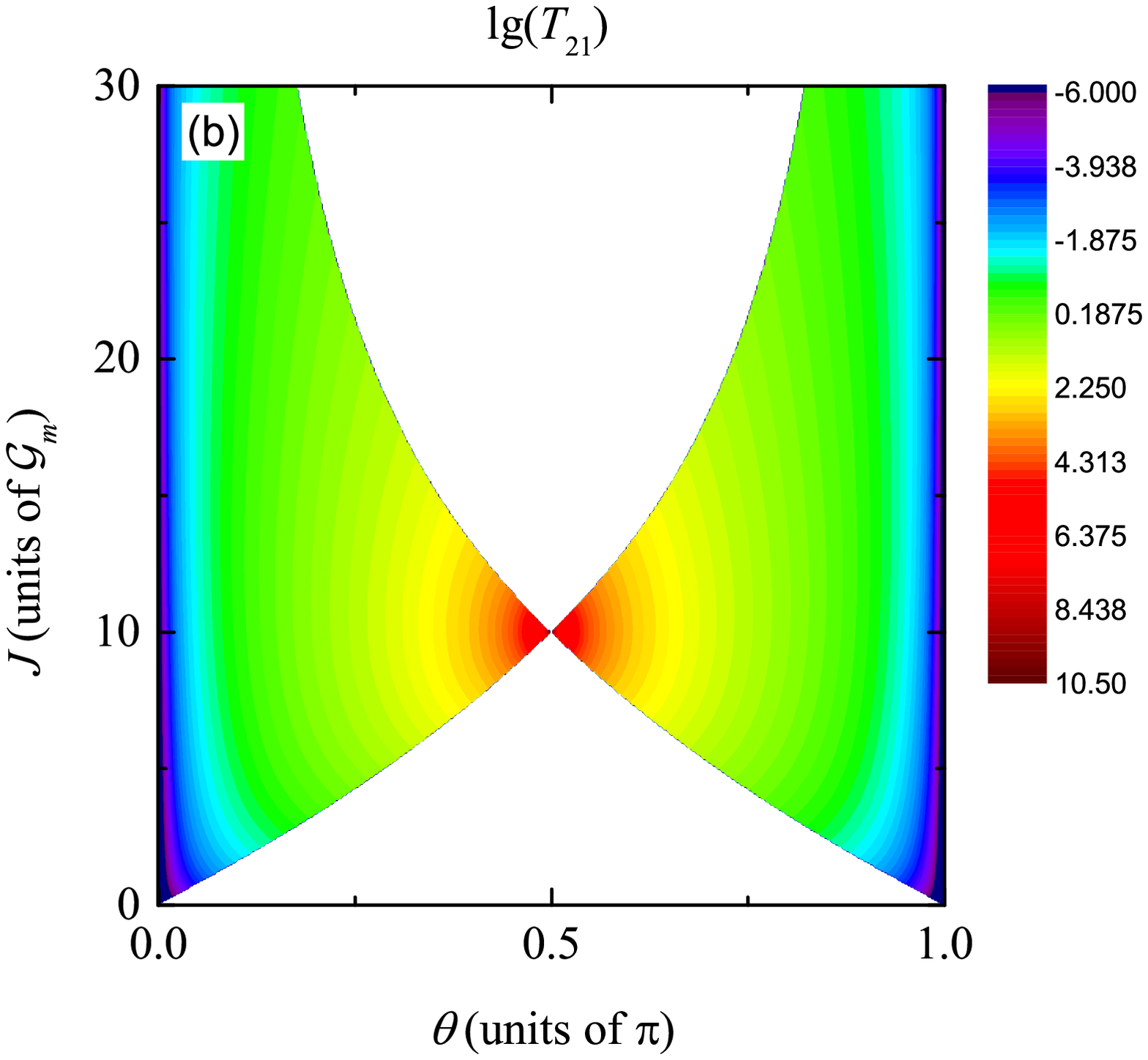}
\caption{(Color online) Contour plot of the logarithm of the transmission
probability $\lg{(T_{21})}$ for (a) $\protect\gamma _{1}=10\mathcal{G}_{m}$,
$\protect\gamma _{2}=15\mathcal{G}_{m}$ and (b) $\protect\gamma _{1}=10%
\mathcal{G}_{m}$, $\protect\gamma _{2}=10\mathcal{G}_{m}$. The optimal
unidirectional conditions are used with $G_{1}=\left\vert G_{2}\right\vert
\equiv G=\protect\sqrt{J\mathcal{G}_{m}/\sin \protect\theta }$ and $\Delta =%
\mathcal{G}_{m}\cos \protect\theta /\sin \protect\theta $. It is shown that
when $\protect\theta \rightarrow \protect\pi /2$, the transmission
probability reaches its maximum. }
\label{fig4}
\end{figure*}

\begin{figure*}[tbp]
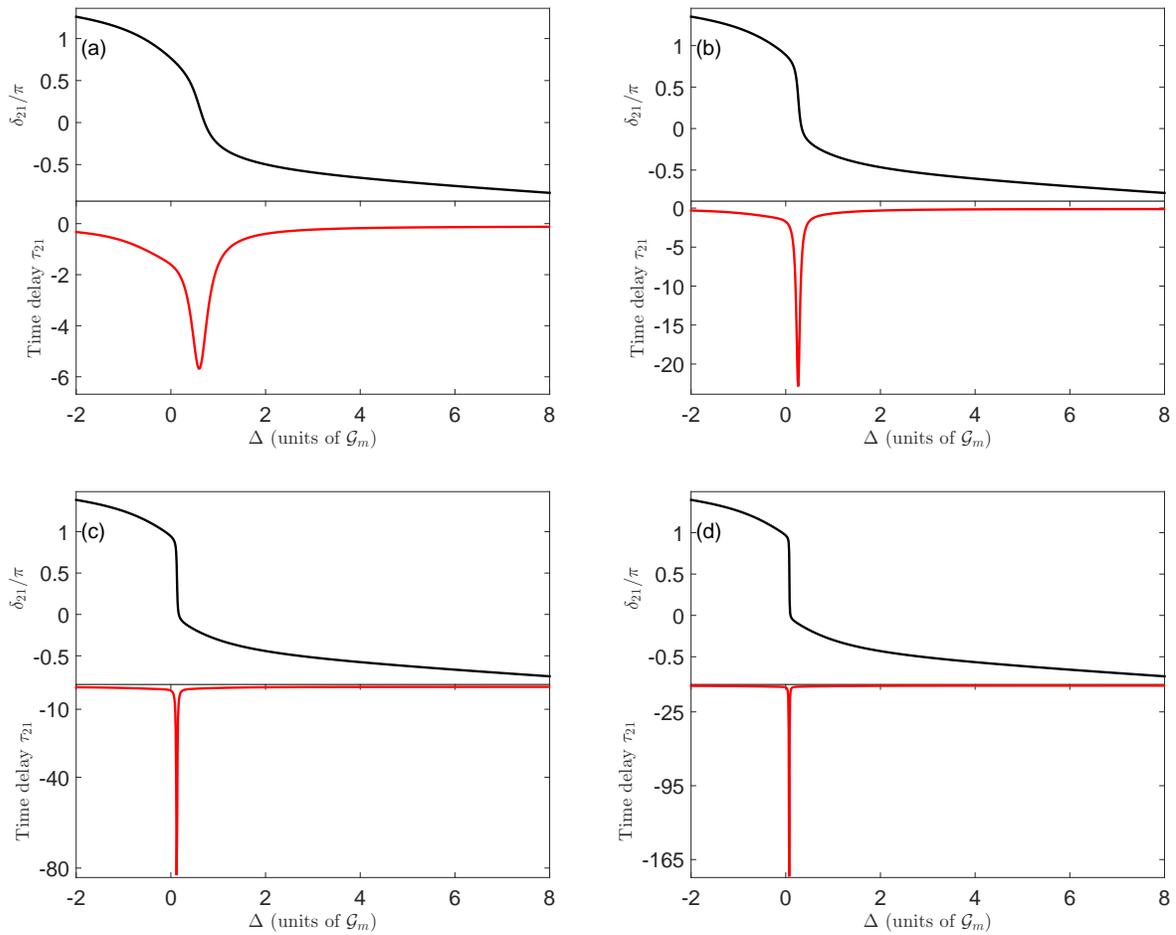

\centering
\includegraphics[bb=0 0 516 409, width=0.45\textwidth,clip]{fig5a.eps} %
\includegraphics[bb=0 0 516 409, width=0.45\textwidth,clip]{fig5b.eps} %
\includegraphics[bb=0 0 516 409, width=0.45\textwidth,clip]{fig5c.eps} %
\includegraphics[bb=0 0 516 409, width=0.45\textwidth,clip]{fig5d.eps}
\caption{(Color online) The group delay $\protect\tau_{21} $ vs the detuning
$\Delta $ for (a) $\protect\theta =0.3\protect\pi $, (b) $\protect\theta =0.4%
\protect\pi $, (c) $\protect\theta =0.45\protect\pi $, and (d) $\protect%
\theta =0.47\protect\pi $. The group delay $\protect\tau_{21} $ is given in
units of $1/\mathcal{G}_{m}$. Other parameters are $J=10\mathcal{G}_{m}$, $%
\protect\gamma _{1}=10\mathcal{G}_{m}$, $\protect\gamma _{2}=15\mathcal{G}%
_{m}$, and $G=\protect\sqrt{J\mathcal{G} _{m}/\sin \protect\theta }$. }
\label{fig5}
\end{figure*}

\subsection{Optomechanical transistor}

\label{ot}

We first analyze the behavior of the transmission matrix elements $t_{12}$
and $t_{21}$, as given by (\ref{t_12}) and (\ref{t_21}). For simplicity, we
assume $G_{1}\equiv G$ with $G>0$, $G_{2}\equiv Ge^{-i\theta }$ with a phase
difference $\theta $ from $G_{1}$, and $\Delta _{i}^{\prime \prime }=\Delta
_{m}\equiv \Delta $ with $\Delta _{i}^{\prime }=\omega _{m}$ for the pump
fields. The transmission coefficients under these conditions can be written
as
\begin{eqnarray}
t_{12} &=&\frac{2\sqrt{\gamma _{1}\gamma _{2}}\left[ -iJ\left( -\mathcal{G}%
_{m}+i\Delta \right) -G^{2}e^{i\theta }\right] }{D_{r}},  \label{t12simp} \\
t_{21} &=&\frac{2\sqrt{\gamma _{1}\gamma _{2}}\left[ -iJ\left( -\mathcal{G}%
_{m}+i\Delta \right) -G^{2}e^{-i\theta }\right] }{D_{r}}  \label{t21simp}
\end{eqnarray}%
with the denominator
\begin{eqnarray}
D_{r} &=&\left( \gamma _{1}+i\Delta \right) \left( \gamma _{2}+i\Delta
\right) \left( -\mathcal{G}_{m}+i\Delta \right)  \notag \\
&&+G^{2}\left( \gamma _{1}+\gamma _{2}+2i\Delta \right) +J^{2}\left( -%
\mathcal{G}_{m}+i\Delta \right)  \notag \\
&&-2iJG^{2}\cos \theta .
\end{eqnarray}%
Using (\ref{t12simp}) for the coefficient $t_{12}$, we choose the phase
difference $\theta $ to satisfy the condition $G=\sqrt{J\mathcal{G}_{m}/\sin
\theta}$ and choose the detuning $\Delta =\mathcal{G}_{m}\cos \theta /\sin
\theta $, which yields that $t_{21}\neq 0$ and $t_{12}=0$, i.e.,
unidirectional propagation of the probe field can be achieved.

We select a set of parameters that satisfy the stability condition using the
result shown in Fig.~\ref{fig2}. Using these parameters, we plot the
logarithms of the transmission probabilities $\lg {(T_{21})}$ and $\lg {%
(T_{12})}$ in Fig.~\ref{fig3}(a) with $T_{ij}=\left\vert t_{ij}\right\vert
^{2}$. The result gives a clear feature of unidirectional amplification of
the probe field in the vicinity of $\Delta =0$, which agrees with our
theoretical prediction. The physics origin of the amplification arises from
the mechanical gain, which can be viewed as a coherent phonon bath that
converts the beam-splitter operation between the mechanical and the cavity
modes to effective parametric processes between these modes. The parametric
processes greatly enhance the photoelastic scattering~\cite{LiuYL}. This
effect is in analogues to the stimulated emission process in atomic systems
when the frequency of the probe field is resonant with that of the
anti-Stokes field, where amplification of the incident photon field can be
achieved. This system can work as an optomechanical transistor at strong
mechanical gain with $\mathcal{G}_{m}\sim \gamma _{1}$ by choosing
appropriate parameters. As shown in Fig.~\ref{fig3}(b), strong
unidirectional amplification can be achieved at $\mathcal{G}_{m}=1.3 \gamma
_{1}$. Meanwhile, the increase of the mechanical gain can induce instability
to this system. With the parameters in Fig.~\ref{fig3}(b), the system
becomes unstable for $\mathcal{G}_{m}/\gamma _{1}>1.325$.

In Fig.~\ref{fig4}, we plot the logarithm of the transmission probability $%
\lg{(T_{21})}$ at the optimal conditions for unidirectional propagation,
i.e., with $G^{2}=J\mathcal{G}_{m}/\sin \theta $ and $\Delta =\mathcal{G}%
_{m}\cos \theta/\sin \theta $. It is shown that in the neighborhood of $%
\theta=\pi /2$, the transmission probability reaches its maximum with $\max{%
(T_{21})}\approx 10^{5}$. 
This strong amplification, together with the nonreciprocity, clearly shows
that our system can be used as an optomechanical transistor facilitated by
the mechanical gain.

\subsection{Ultra-long optical delay}

\label{ol}

The optical group delay is another important parameter to characterize the
optical transmission and responses. It is well known that the optical
transmission within an electromagnetically-induced transparency window
experiences a dramatic reduction in its group velocity. Similar effects can
be expected in the optical transmission in optomechanical systems. Here we
investigate the optical delay in our system. We first introduce the optical
group delay time defined in terms of the phase of the transmitted probe
field as
\begin{equation}
\tau _{ij}=\frac{d\delta _{ij}}{d\omega _{p}},
\end{equation}%
where $\delta _{ij}=\arg \left[ t_{ij}\left( \omega _{p}\right) \right] $ is
the phase of the output field at the frequency $\omega _{p}$~\cite%
{OMIT2,OMIT4}. We consider the system operated in the regime of an
optomechanical transistor with $|t_{21}|\gg 1$ and $t_{12}=0$. To ensure
unidirectional amplification and similar to the previous subsection, we let
the parameters satisfy the relations: $J=10\mathcal{G}_{m}$, $G_{1}\equiv G$
$\left( G>0\right) $, $G_{2}\equiv Ge^{-i\theta }$, $\Delta _{i}^{\prime
\prime }\equiv \Delta $, $\gamma _{1}=10\mathcal{G}_{m}$, $\gamma _{2}=15%
\mathcal{G}_{m}$, and $G^{2}=J\mathcal{G}_{m}/\sin \theta $. In Fig.~\ref%
{fig5}, we plot the phase $\delta _{21}$ and the group delay $\tau_{21} $ as
functions of the detuning $\Delta $. It can be shown that strong group delay
occurs in the working window of the optomechanical transistor. As $\theta $
approaches to the value of $\pi /2$, the group delay exhibits sharp
increase. This indicates that the strengthening of the amplification process
gives rise to dramatic increase in the group delay. Note that near $\theta
=\pi /2$, as shown in Fig.~\ref{fig2}(a), the system is close to the
boundary between the stable and the unstable regions, and is more fragile to
environmental disturbance. Therefore, there is a tradeoff between the
amplification and group delay and the stability of this system. By selecting
appropriate parameters, one can realize an optomechanical transistor with
significant time delay.

\section{Conclusions}

\label{conclusions}

To conclude, we have shown that an optomechanical transistor can be realized
in a cyclic optomechanical system with finite mechanical gain. The
nonreciprocal behavior of this system arises from the phase difference
between the optomechanical couplings $G_{1}$ and $G_{2}$, which breaks the
time-reversal symmetry of this system. The mechanical gain does not affect
the nonreciprocity of the optical response, but it plays a key role in
achieving the amplification for the probe field. The presence of the
mechanical gain can generate strong parametric processes between the
mechanical mode and the cavities and significantly enhance the photoelastic
scattering when the probe field is at an optimal frequency. Combining the
phase difference between the couplings with the mechanical gain thus enables
the unidirectional amplification of the probe field. Furthermore, the
amplification of the probe field is accompanied by an ultra-long group delay
in the output field. Our work hence provides an effective approach to
control the light propagation in an optomechanical system and could
stimulate future studies of nonreciprocal optomechanical interfaces in
nonlinear photonic devices.

\acknowledgments This work is supported by the National Key R\&D Program of
China grant 2016YFA0301200, the National Basic Research Program (973
Program) of China (under Grant No.~2014CB921403), the National Natural
Science Foundation of China (under Grants No.~11505126, No.~11774024,
No.~11534002, No.~U1530401), and the Postdoctoral Science Foundation of
China (under Grant No. 2016M591055). X.Z.Z. is supported by PhD research
startup foundation of Tianjin Normal University under Grant No. 52XB1415.
L.T. is supported by the National Science Foundation (USA) under Award
Numbers DMR-0956064 and PHY-1720501.

\end{document}